\documentclass[a4paper,fleqn]{cas-sc}
\usepackage{color,array,amsthm}
\usepackage{graphicx}
\usepackage{cite}
\usepackage{amsmath,amssymb,amsfonts}
\usepackage{algorithmic}
\usepackage{algorithm}
\usepackage{graphicx}
\usepackage{textcomp}
\usepackage{xcolor}
\usepackage{gensymb}
\usepackage{multirow}
\usepackage{amsmath}
\usepackage[authoryear]{natbib}

\def\tsc#1{\csdef{#1}{\textsc{\lowercase{#1}}\xspace}}
\tsc{WGM}
\tsc{QE}
\tsc{EP}
\tsc{PMS}
\tsc{BEC}
\tsc{DE}
\begin{document}
\let\WriteBookmarks\relax
\def\floatpagepagefraction{1}
\def\textpagefraction{.001}
\shorttitle{Assessing Ionospheric Scintillation Risk for D2C SatComs using Frequency-Scaled GNSS Observations}
\shortauthors{A.M. Darya et~al.}

\title[mode = title]{Assessing Ionospheric Scintillation Risk for Direct-to-Cellular Satellite Communications using Frequency-Scaled GNSS Observations} 

\author[1]{Abdollah Masoud Darya}
\cormark[1]
\ead{abdollah.masoud@ieee.org}

\author[1,2]{Muhammad Mubasshir Shaikh}
\ead{mshaikh@sharjah.ac.ae}

\affiliation[1]{organization={SSAH, University of Sharjah},
                city={Sharjah 27272},
                country={United Arab Emirates}}
\affiliation[2]{organization={Electrical Engineering Dept., University of Sharjah},
                city={Sharjah 27272},
                country={United Arab Emirates}}

\cortext[cor1]{Corresponding author}

% \author{Abdollah Masoud Darya}
% \member{Graduate Student Member, IEEE}
% \affil{SAASST, University of Sharjah, Sharjah, UAE} 

% \author{Muhammad Mubasshir Shaikh}
% \affil{Electrical Engineering Dept., University of Sharjah, Sharjah, UAE\\SAASST, University of Sharjah, Sharjah, UAE} 

% \receiveddate{Manuscript received XXXXX 00, 0000; revised XXXXX 00, 0000; accepted XXXXX 00, 0000.}
%% \accepteddate{XXXXX XX XXXX}
%% \publisheddate{XXXXX XX XXXX}

% \corresp{{\itshape(Corresponding author: Abdollah Masoud Darya)}}

% \authoraddress{Abdollah Masoud Darya is with SAASST, University of Sharjah, Sharjah, UAE (e-mail: abdollah.masoud@ieee.org). Muhammad Mubasshir Shaikh is with the Electrical Engineering Department, College of Engineering, University of Sharjah, Sharjah, UAE (e-mail: mshaikh@sharjah.ac.ae) and with SAASST, University of Sharjah, Sharjah, UAE.}

%\editor{Mentions of supplemental materials and animal/human rights statements can be included here.}
%\supplementary{Color versions of one or more of the figures in this article are available online at {http://ieeexplore.ieee.org}.}

% \markboth{CORRESPONDENCE}{}

\begin{abstract}
One of the key issues facing Direct-to-Cellular (D2C) satellite communication systems is ionospheric scintillation on the uplink and downlink, which can significantly degrade link quality. This work investigates the spatial and temporal characteristics of amplitude scintillation at D2C frequencies by scaling L-band scintillation observations from Global Navigation Satellite Systems (GNSS) receivers to bands relevant to D2C operation, including the low-band, and 3GPP's N255 and N256. These observations are then compared to scaled radio-occultation scintillation observations from the FORMOSAT-7/COSMIC-2 (F7/C2) mission, which can be used in regions that do not possess ground-based scintillation monitoring stations. As a proof of concept, five years of ground-based GNSS scintillation data from Sharjah, United Arab Emirates, together with two years of F7/C2 observations over the same region, corresponding to the ascending phase of Solar Cycle 25, are analyzed. Both space-based and ground-based observations indicate a pronounced diurnal scintillation peak between 20–22 local time, particularly during the equinoxes, with occurrence rates increasing with solar activity. Ground-based observations also reveal a strong azimuth dependence, with most scintillation events occurring on southward satellite links. The scintillation occurrence rate at the low-band is more than twice that observed at N255 and N256, highlighting the increased robustness of higher D2C bands to ionospheric scintillation. These results demonstrate how GNSS scintillation observations can be leveraged to characterize and anticipate scintillation-induced D2C link impairments, which help in D2C system design and the implementation of scintillation mitigation strategies.
\end{abstract}


\maketitle

\section{Introduction}
One of the key pillars of the sixth generation (6G) communications paradigm is ubiquitous connectivity \citep{jamshed2025non}. The idea of being connected everywhere at all times is difficult to realize by terrestrial-bound base-stations. Yet, the recent emergence of the \emph{Direct-to-Cellular} (D2C) approach, where standard cellular phones can connect directly to satellites, and vice versa, aims to address this issue \citep{andrews20246g}.\par
As satellite signals propagate between the transmitter and ground receiver, they are affected by natural propagation effects or artificial sources of interference that reduce the quality of service. The fluctuation of the amplitude and phase of radio waves (\emph{scintillation}) due to the signal's propagation through the ionosphere is one of these sources \citep{Zhao2026Global}. Scintillations can occur naturally through interactions between the signal and ionospheric electron-density irregularities \citep{Wernik2003Ionospheric}, or they can be caused by external interfering signals \citep{pica2023analysis}. Ionospheric scintillation significantly affects signals under $3\,\mathrm{GHz}$ \citep{itu}, and its effects vary based on several spatial and temporal parameters, such as geographical location, time of day, season, and solar cycle progression.\par
Global Navigation Satellite Systems (GNSS)-based applications have long faced the issue of scintillation, and their effects on accuracy, availability, and integrity, especially at equatorial and high latitude regions \citep{aguiar2025impact}. Even so, the effect on D2C communications systems would be more prominent. This is due to the two-way nature of D2C, with standard users being a part of both uplink and downlink, compared to only downlink for standard GNSS users. Additionally, GNSS is broadcast-based and is resilient to short-term temporal fluctuations due to the usage of tracking techniques \citep{vila2018mitigation}. However, D2C communications, especially voice or broadband connectivity, are sensitive to abrupt signal fluctuations.\par

To better understand the risk of ionospheric scintillation on D2C systems, dedicated scintillation monitoring stations must be installed and long-term observation campaigns conducted to capture scintillation variations over the 11-year solar cycle. In practice, however, establishing and maintaining such infrastructure over a full solar cycle is resource-intensive and cannot provide timely insight for the D2C deployments already underway \citep{garcia2025direct}. Although companies deploying D2C systems may continue developing dedicated monitoring networks and long-term campaigns, alternative approaches are needed in the meantime. This work is motivated by the need for such an alternative approach.\par

Ground-based GNSS ionospheric scintillation monitoring stations have long been established to monitor ionospheric irregularities and their impact on GNSS signals \citep{dodson2001ionospheric,jayachandran2009canadian,vani2021monitoring}. Furthermore, radio-occultation (RO) experiments onboard missions such as FORMOSAT-7/COSMIC-2 (F7/C2) \citep{yue2014space} extend the coverage of observations to include regions with no established scintillation monitoring networks. This work proposes scaling widely available GNSS L-band amplitude-scintillation observations to D2C frequencies to assess scintillation risk to D2C links as a function of local time, season, signal arrival direction, and solar activity. Here, we provide a proof of concept on how this can be achieved by using data from a multi-frequency GNSS reference receiver located at Sharjah, United Arab Emirates. Furthermore, the ground-based observations are compared to space-based RO amplitude scintillation observations from F7/C2 to gauge the suitability of using space-based RO observations in regions that lack ground-based scintillation monitoring receivers.\par

In the literature, the investigation of multi-frequency ionospheric scintillation has recently been proposed by \cite{Xu2026Multi} using an observation-based approach, and by \cite{Sun2026Ionospheric} with a simulation-based approach. Furthermore, the scaling of ionospheric scintillation from one frequency to another was investigated previously. For instance, L-band scintillation was scaled to very high frequency (VHF) in \cite{bhattacharyya2019signal}, and the different GNSS L-band frequencies (L1/L2/L5) in \cite{song2021multifrequency} and \cite{carrano2014inverse}. Additionally, using GNSS L-band scintillation observations as a proxy model for scintillation at UHF was presented in \cite{caton2004gps}. Yet, what is missing from the literature is a self-contained long-term study on the scaling of scintillation from L-band to D2C frequencies, including comparisons with space-based RO measurements, and the conclusions that can be derived as a result.\par

\section{Methodology}

\begin{figure}
\centering
\includegraphics[width=0.6\linewidth]{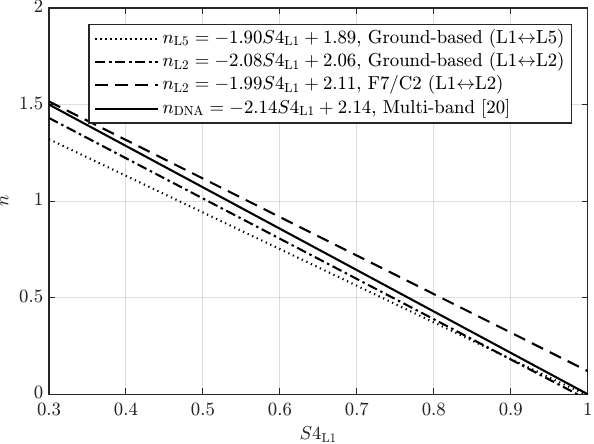}
\caption{\small{frequency exponent $n$ versus $S4_\mathrm{L1}$.}}
\label{Fig1}
\end{figure}

\begin{table}
\centering
\caption{Performance metrics of the LS best-fit and DNA}
\label{Tab1}
\begin{tabular}{|l|cc|cc|}
\hline
\multirow{2}{*}{Scaling} & \multicolumn{2}{c|}{LS best-fit} & \multicolumn{2}{c|}{DNA} \\ \cline{2-5} 
 & \multicolumn{1}{c|}{RMSE} & \multicolumn{1}{c|}{$\mathrm{R}^{2}$} & \multicolumn{1}{c|}{RMSE} & \multicolumn{1}{c|}{$\mathrm{R}^{2}$} \\ \hline
L1 to L5 & \multicolumn{1}{l|}{0.085} & 0.76 & \multicolumn{1}{l|}{0.086} & 0.76\\ \hline
L1 to L2 (Ground-based) & \multicolumn{1}{l|}{0.074} & 0.82 & \multicolumn{1}{l|}{0.074} & 0.82\\ \hline
L1 to L2 (F7/C2)& \multicolumn{1}{l|}{0.087} & 0.72 & \multicolumn{1}{l|}{0.085} & 0.73\\ \hline
\end{tabular}
\end{table}

In this work, we utilize GNSS Ionospheric scintillation observations to assess how D2C signals might be affected. We rely on a Septentrio PolaRx5S multi-frequency multi-constellation reference GNSS receiver located at Sharjah (Geographic Latitude: $25.28\degree$, Geographic Longitude: $55.46\degree$), United Arab Emirates, which provides data over the ascending phase of solar cycle $25$ ($2020$--$2024$). The retrieved observations correspond to L1 scintillation from GPS L1 C/A and Galileo E1, both having a carrier frequency of $1\,575.42\,\mathrm{MHz}$. Additionally, to limit the influence of multipath on scintillation, this work utilizes observations with an elevation greater than $30\degree$. Furthermore, to remove the influence of geomagnetically active periods, observations that coincide with either a planetary K-index (Kp) greater than $4$ or a disturbance storm-time (DST) index less than $-50\,\mathrm{nT}$ were removed.\par
This work also utilizes RO amplitude scintillation observations from F7/C2 for the years $2023$--$2024$\footnote{The first available scintillation observation from F7/C2 was in $25/03/2022$. Observations of the year $2022$ were omitted to focus on complete years.}. The longitude and latitude values of the observations were limited to $54\degree\leq\text{longitude}\leq57\degree$ and $23\degree\leq\text{latitude}\leq27\degree$ to enable a comparison with observations from the ground-based receiver by matching the region of coverage. Additionally, only GPS observations were used in this work for F7/C2.\par 
Amplitude scintillation is represented by the standard deviation of the signal's intensity (I) normalized to the average signal intensity over $60$ seconds \citep{3gpp}, i.e.,
\begin{equation}\label{Eq1}
    S4=\sqrt{\frac{\langle I^2\rangle-\langle I\rangle^2}{\langle I\rangle^2}}.
\end{equation}
The severity of scintillation can be inferred from the magnitude of $S_{4}$, where weak, moderate, and strong scintillation are represented by $S4<0.3$, $0.3\leq S4\leq0.6$, and $S4>0.6$, respectively \citep{3gpp}. This work focuses on strong scintillation, since such events can cause peak-to-peak power fluctuations greater than  $14\,\mathrm{dB}$ \citep{itu}.\par
To translate the scintillation occurrences seen at L-band GNSS signals to those that could be seen at D2C signals, we use the standard frequency scaling relation \citep{3gpp}
\begin{equation}\label{Eq2}
    S4_\mathrm{f2}=S4_\mathrm{f1}\left(\frac{f_2}{f_1}\right)^{-n},
\end{equation}
where $S4_{f_1}$ is the S4 observation at frequency $f_1$, and $S4_{f_2}$ is the scaled S4 at frequency $f_2$. Furthermore, the frequency exponent $n$, which characterizes the signal frequency dependence of amplitude scintillation \citep{bhattacharyya2019signal}, is represented by
\begin{equation}\label{Eq3}
n=\frac{\ln \left(\frac{S4_\mathrm{f2}}{S4_\mathrm{f1}}\right)}{\ln \left(\frac{f_1}{f_2}\right)}.
\end{equation}
The frequency scaling exponent $n$ for frequencies ranging from $30\,\mathrm{MHz}$ to $6\,\mathrm{GHz}$, as observed from multiple satellite measurements, was found to range between $1$ to $2$ (see Table 3.4 in \cite{wheelon2003electromagnetic}). Of these different experiments, the observations taken by the Defense Nuclear Agency (DNA) wideband satellite experiment were the most prolific. The DNA satellite acted as a multi-frequency coherent radio beacon targeted towards studying the ionospheric scintillation of signals in the VHF, UHF, L, and S-bands. Through its observations, the value $n=1.5$ was empirically derived \citep{fremouw1978early} for the weak scattering assumption, i.e., weak to moderate scintillation \citep{3gpp,wheelon2003electromagnetic}. As the magnitude of S4 increases and approaches unity, $n$ saturates at $0$ \citep{3gpp,fremouw1978early}. Therefore, the trend of $n$ can be well represented by a linear relationship between moderate and strong values of scintillation, where $n=1.5$ at $S4=0.3$ and $n=0$ at $S4=1$ \citep{bhattacharyya2019signal}.\par
To see if this DNA-derived linear trend is applicable in our scenario, five years of simultaneous observations of GPS-derived $S4_\mathrm{L1}$/$S4_\mathrm{L2}$/$S4_\mathrm{L5}$ from the ground-based receiver and two years of simultaneous observations of $S4_\mathrm{L1}$/$S4_\mathrm{L2}$ from F7/C2 were used, with the value of $n$ calculated as in \eqref{Eq3} for $0.3\leq S4\leq1$. Next, a least-squares approach was utilized to obtain the best-fit line for $n_\mathrm{L2}$ that scales $S4_\mathrm{L1}$ to $S4_\mathrm{L2}$, and $n_\mathrm{L5}$ that scales $S4_\mathrm{L1}$ to $S4_\mathrm{L5}$. The least-squares best fit equations for each case are presented in Fig. \ref{Fig1}. Note that the reverse procedure, e.g., scintillation observed at L5 scaled to L1 instead, would yield similar results to L1 scintillation scaled to L5.\par
In Fig. \ref{Fig1}, the solid line represents the $n_\mathrm{DNA}$ proposed by \cite{fremouw1978early}, and as can be seen it is very close in magnitude to our empirically derived ground-based and F7/C2 $n_\mathrm{L2}$ and $n_\mathrm{L5}$, indicating the suitability of the DNA-derived $n_\mathrm{DNA}$ values. Since $n_\mathrm{DNA}$ was derived from multi-band observations, it will be used in this work to scale S4 from GNSS to D2C frequencies. To evaluate the performance consequences of this choice, the $n_\mathrm{L2}$ and $n_\mathrm{L5}$ were used to scale scintillation observed on L1 to L2 and L5, respectively. Then, $n_\mathrm{DNA}$ was used to perform the same procedure. The scaled scintillation values were then compared to the true observed L2/L5 values, and the errors in terms of RMSE and $\mathrm{R}^{2}$ are provided in Table \ref{Tab1}. As can be seen from Table \ref{Tab1}, the performance metrics almost match, which proves the suitability of using $n_\mathrm{DNA}$ to scale scintillation from GNSS to D2C frequencies.\par

\begin{table}
\centering
\caption{Bands of D2C signals}
\label{Tab2}
\begin{tabular}{|l|c|c|c|}
\hline
Band & \begin{tabular}[c]{@{}l@{}}Uplink Freq. (MHz)\end{tabular} & \begin{tabular}[c]{@{}l@{}}Downlink Freq. (MHz)\end{tabular} & Reference \\ \hline
\begin{tabular}[c]{@{}l@{}}Low-band (AST and AT\&T)\end{tabular} & $698$--$849$ & $728$--$894$ & \citep{FccAST} \\ \hline
\begin{tabular}[c]{@{}l@{}}PCS G Block (Starlink and T-Mobile)\end{tabular} & $1\,910$--$1\,915$ & $1\,990$--$1\,995$ & \citep{FccStarlink} \\ \hline
NR N255 & $1\,626$--$1\,660$ & $1\,525$--$1\,559$ & \citep{3gppV2} \\ \hline
NR N256 & $1\,980$--$2\,010$ & $2\,170$--$2\,200$ & \citep{3gppV2} \\ \hline
\end{tabular}
\end{table}

\begin{figure}
\centering
\includegraphics[width=0.6\linewidth]{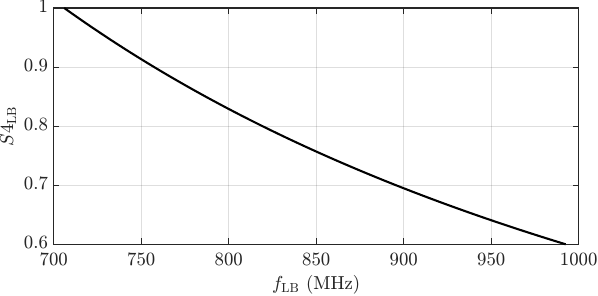}
\caption{\small{$\mathrm{S4}_\mathrm{LB}$ values corresponding to $S4_\mathrm{L1}=0.3$.}}
\label{Fig2}
\end{figure}

D2C communication would be performed in two ways. Spectrum sharing between telecommunication and satellite companies, such as AST SpaceMobile with AT\&T or Starlink and T-Mobile, or using new spectrum such as NR N255 and N256 introduced by the 3GPP Standard Release 17 (see Table \ref{Tab2} for details). Due to the frequency dependence of scintillation, the low-band (LB) is particularly prone to ionospheric scintillation as can be seen in Fig. \ref{Fig2}, where moderate scintillation seen in L1 ($S4=0.3$) equals strong scintillation ($S4>0.6$) in the LB.\par
This work will focus on three bands by scaling the L1 scintillation to three distinct frequencies: N255 represented by $1\,600\,\mathrm{MHz}$, N256 represented by $2\,000\,\mathrm{MHz}$ (this also covers PCS G Block due to similar frequency values), and LB represented by $800\,\mathrm{MHz}$.\par

\section{Observations}

\begin{table}
\centering
\caption{Number of strong scintillation observations}
\label{Tab3}
\begin{tabular}{|l|cc|}
\hline
\multirow{2}{*}{Band} & \multicolumn{2}{l|}{Number of $S4>0.6$ observations} \\ \cline{2-3} 
 & \multicolumn{1}{l|}{Ground-based} & F7/C2 \\\hline
L1 & \multicolumn{1}{c|}{$\phantom{0}9\,652$} & $1\,460$\\\hline
LB & \multicolumn{1}{c|}{$25\,875$} & $5\,275$\\\hline
NR N255 & \multicolumn{1}{c|}{$\phantom{0}9\,447$} & $1\,409$\\\hline
NR N256 & \multicolumn{1}{c|}{$\phantom{0}7\,432$} & $\phantom{0}\,982$\\\hline
\end{tabular}
\end{table}

\begin{figure}
\centering
\includegraphics[width=0.6\linewidth]{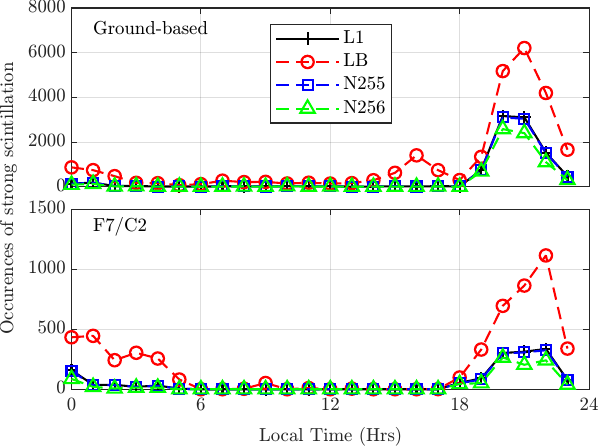}
\caption{\small{Hourly trend of strong scintillation occurrences.}}
\label{Fig3}
\end{figure}

\begin{figure}
\centering
\includegraphics[width=0.6\linewidth]{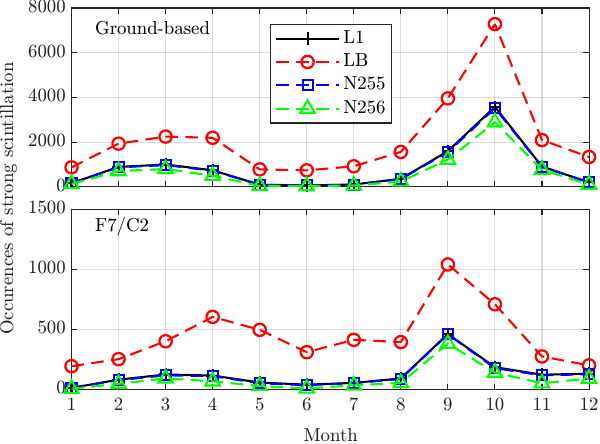}
\caption{\small{Monthly trend of strong scintillation occurrences.}}
\label{Fig4}
\end{figure}

\begin{figure}
\centering
\includegraphics[width=0.6\linewidth]{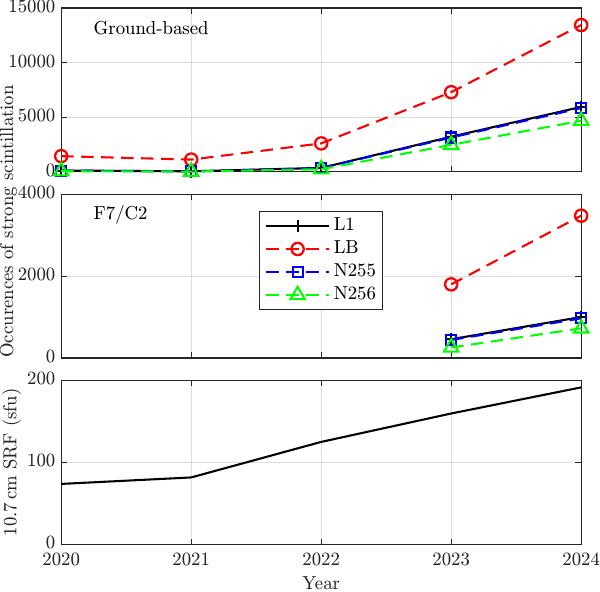}
\caption{\small{Yearly trend of strong scintillation occurrences. Lower plot: F10.7 solar radio flux in sfu.}}
\label{Fig5}
\end{figure}

\begin{figure}
\centering
\includegraphics[width=0.6\linewidth]{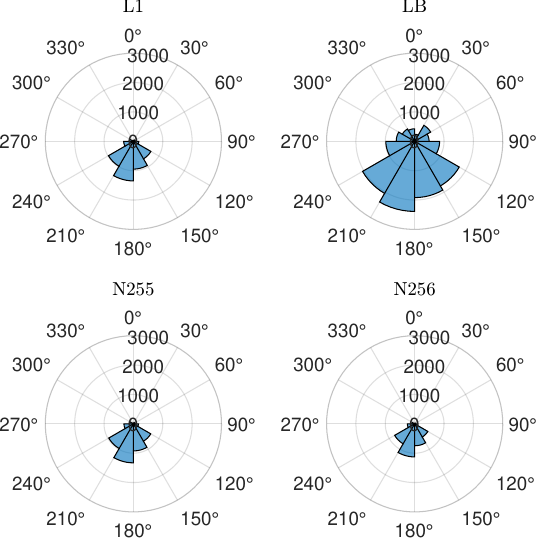}
\caption{\small{Azimuth of strong scintillation occurrences.}}
\label{Fig6}
\end{figure}

After scaling the L1 scintillation to the three distinct frequencies, Figs. \ref{Fig3}--\ref{Fig6} were plotted to assess the spatial and temporal characteristics of scintillation at D2C frequencies. These trends represent the number of minute-by-minute occurrences (or observations) of strong scintillation.\par
The number of ground-based strong scintillation observations (Table \ref{Tab3}) was $6.61\times$ times higher than the F7/C2 observations. This is simply due to the higher cadence of ground-based observations. Over the entire study period, the number of scintillation observations in the LB was $2.68\times$ and $3.61\times$ those observed at L1 from ground-based and F7/C2 measurements, respectively. In contrast, the occurrence rates at N255 were comparable to L1, at approximately $0.98\times$ and $0.97\times$ of the L1 ground-based and F7/C2 observations. For N256, the corresponding occurrences were lower, at $0.77\times$ and $0.67\times$ of the L1 ground-based and F7/C2 observations, respectively. This negative frequency dependence of scintillation, where lower frequencies are more susceptible to strong scintillation, was observed previously by \cite{Xu2026Multi}.\par
Fig. \ref{Fig3} shows the diurnal variation of scintillation. The solid line represents the L1 strong scintillation observations, while the circle markers represent the LB observations, followed by N255 and N256, represented by the square and triangle markers respectively.\par
The largest peak of scintillation as observed by F7/C2 and the ground-based receiver is seen at $20$--$22$ local time for all bands, corresponding to the post-sunset to midnight period. A second minor peak is also seen by the ground-based receiver at the LB at $15$--$17$ local time which represents the daytime pre-sunset period. This secondary peak is typically observed as weak L-band scintillation \citep{Shaikh2021Daytime,Shaikh2020Occurrence,darya2024multi}, and is typically associated with the sporadic E layers \citep{Shaikh2024Characteristics}, but due to the frequency dependence of scintillation this weak scintillation at L-band is translated to strong scintillation in the LB \citep{caton2004gps}. Another peak is observed after midnight by both instruments, with F7/C2 observing it for a more extended period up to 6 hours.\par
The number of strong scintillation occurrences at the LB at its peak is more than double the peak of N255 and N256 as observed by ground and space-based instruments, which highlights the challenge to be faced by LB D2C systems. Fig. \ref{Fig3} also shows that even N256 with its higher frequency will be subjected to considerable levels of strong scintillation.\par

Fig. \ref{Fig4} shows the seasonal variation of scintillation. For all trends, the summer and winter seasons saw the lowest number of scintillation observations, while the equinoxes, and particularly the autumnal equinox (September--October), had the highest occurrence of scintillation, for both ground and space-based instruments. This equinoctial enhancement of scintillation was also observed by \cite{Xu2026Multi}. This seasonal enhancement is commonly attributed to the Russell–McPherron effect \citep{russell1973semiannual}, enabled due to the reverse alignment between the Earth's magnetic field and that of the solar wind \citep{buschmann2024statistical}. Similar to Fig. \ref{Fig3}, the occurrence peaks for LB during the equinoxes are more than twice that for N255 and N256.\par

Fig. \ref{Fig5} shows the yearly variation of scintillation, which shows the impact of solar activity. Additionally, Fig. \ref{Fig5} (lower) shows the solar radio flux at 10.7 cm in solar flux units (sfu)\footnote{The data was obtained from the GSFC/SPDF OMNIWeb interface at https://omniweb.gsfc.nasa.gov}, which is used as an indicator for solar activity. The sun undergoes an $11$ year cycle of activity, and the peak of the current solar cycle (cycle $25$) was in $2024$ \citep{jha2024predicting}. Therefore, observing the increase in strong scintillation from the solar cycle minimum in $2020$ to its peak in $2024$ allows us to understand how solar activity impacts the number of strong scintillation events at different D2C frequencies.\par
Each trend shows similar values in $2020$ and $2021$ followed by a gradual increase in $2022$ and a sharp increase in $2023$ and $2024$. This increase is also reflected in the F7/C2 observations. Similar to Figs. \ref{Fig3}--\ref{Fig4}, the number of LB strong scintillation occurrences is more than twice the other bands.\par

Fig. \ref{Fig6} shows the spatial distribution of scintillation, with each subplot representing a different band. The number of strong scintillation occurrences is represented by the radial axis, while the azimuth is represented by the angular axis with a resolution of $30\degree$.\par
All subplots show that the majority of strong scintillation is seen from the south. In fact, the percentage of northerly occurrences was $23\%$ for LB, and $7\%$ and $5\%$ for N255 and N256, respectively. This shows that almost all strong scintillation is induced on N255 and N256 links from the south, while almost three-quarters of all strong observations on LB are from the south. This is due to scintillation effects being more intense south of the receiver location which falls within the northern equatorial ionization anomaly crest, as compared to mid-latitude regions north of the receiver \citep{dunn2024temporal}.\par 
Studying the spatial and temporal characteristics of ionospheric scintillation at D2C frequencies for different regions is essential for assessing link robustness and planning operational strategies. For example, temporal patterns help identify periods in which the user-satellite links are more likely to degrade, enabling operators to schedule adaptive mitigation techniques \citep{vila2018mitigation}. Spatial patterns, particularly the strong southerly azimuth dependence observed in this region, enable operators to enact strategies that prioritize geometries less affected by scintillation. For example, since D2C constellations are large, higher weights can be assigned to northern communication links during periods of anticipated strong scintillation. Having these insights allows D2C system operators to anticipate performance degradation rather than react to it, which improves the quality of service of D2C systems.\par

\section{Conclusion}
This work demonstrated the use of scaled GNSS-frequency scintillation observations to characterize the spatial and temporal variability of amplitude scintillation at D2C frequencies. As a case study, five years of ground-based data from a GNSS receiver in Sharjah, United Arab Emirates, together with two years of space-based observations from the F7/C2 mission over the same region, were analyzed. Both datasets revealed a pronounced diurnal peak between 20–22 local time, with scintillation occurrences concentrated around the equinoxes and increasing with solar activity. Ground-based observations further indicated a clear azimuth dependence, with the majority of strong scintillation occurrences associated with southward satellite links. These results demonstrate that scaled GNSS scintillation observations can serve as a key indicator of scintillation risk for D2C link reliability. Future work will focus on validating these findings using measurements from D2C ground-based monitoring stations and satellite-derived signal quality metrics.

\section*{Acknowledgments}
This research was supported in part by the University of Sharjah grant number $26020403372$.

\bibliographystyle{cas-model2-names}
\bibliography{References.bib}

@article{jamshed2025non,
author={Jamshed, Muhammad Ali and Kaushik, Aryan and Dajer, Miguel and Guidotti, Alessandro and Parzysz, Fanny and Lagunas, Eva and Di Renzo, Marco and Chatzinotas, Symeon and Dobre, Octavia A.},
  journal={IEEE Communications Standards Magazine}, 
  title={Non-Terrestrial Networks for {6G}: Integrated, Intelligent, and Ubiquitous Connectivity}, 
  year={2025},
  volume={9},
  number={3},
  pages={86--93}
}

@article{Wernik2003Ionospheric,
title = {Ionospheric irregularities and scintillation},
journal = {Advances in Space Research},
volume = {31},
number = {4},
pages = {971-981},
year = {2003},
issn = {0273-1177},
author = {A.W. Wernik and J.A. Secan and E.J. Fremouw}
}

@article{pica2023analysis,
  author={Pica, Emanuele and Minetto, Alex and Cesaroni, Claudio and Dovis, Fabio},
  journal={IEEE Journal of Selected Topics in Applied Earth Observations and Remote Sensing}, 
  title={Analysis and Characterization of an Unclassified {RFI} Affecting Ionospheric Amplitude Scintillation Index Over the Mediterranean Area}, 
  year={2023},
  volume={16},
  number={},
  pages={8230--8248}
}

@article{vila2018mitigation,
  author={Vilà-Valls, Jordi and Closas, Pau and Fernández-Prades, Carles and Curran, James Thomas},
  journal={IEEE Transactions on Aerospace and Electronic Systems}, 
  title={On the Mitigation of Ionospheric Scintillation in Advanced {GNSS} Receivers}, 
  year={2018},
  volume={54},
  number={4},
  pages={1692--1708}
}

@article{aguiar2025impact,
author = {Aguiar, C. R. and Monico, J. F. G. and Moraes, A. O.},
title = {Impact of Ionospheric Scintillations on {GNSS} Availability and Precise Positioning},
journal = {Space Weather},
volume = {23},
number = {2},
pages = {e2024SW004217},
year = {2025}
}

@techreport{3gpp,
 author = {3GPP},
 institution = {3rd Generation Partnership Project (3GPP)},
 month = {09},
 note = {Version 15.4.0},
 number = {38.811},
 title = {Study on {New Radio (NR)} to support non-terrestrial networks},
 year = {2020}
}

@article{bhattacharyya2019signal,
author = {Bhattacharyya, A. and Gurram, P. and Kakad, B. and Sripathi, S. and Sunda, S.},
title = {Signal Frequency Dependence of Ionospheric Scintillations: An Indicator of Irregularity Spectrum Characteristics},
journal = {Journal of Geophysical Research: Space Physics},
volume = {124},
number = {10},
pages = {8081--8091},
year = {2019}
}

@article{song2021multifrequency,
  author={Song, Kaili and Meziane, Karim and Kashcheyev, Anton and Jayachandran, P. T.},
  journal={IEEE Transactions on Geoscience and Remote Sensing}, 
  title={Multifrequency Observation of High Latitude Scintillation: A Comparison With the Phase Screen Model}, 
  year={2022},
  volume={60},
  number={},
  pages={1--9},
}

@inproceedings{carrano2014inverse,
  title={An inverse diffraction technique for scaling measurements of ionospheric scintillations on the {GPS L1, L2, and L5} carriers to other frequencies},
  author={Carrano, Charles S and Groves, Keith M and Delay, Susan H and Doherty, Patricia H},
  booktitle={Proceedings of the 2014 international technical meeting of the institute of navigation},
  pages={709--719},
  year={2014}
}

@article{jayachandran2009canadian,
  author={Jayachandran, P. T. and Langley, R. B. and MacDougall, J. W. and Mushini, S. C. and Pokhotelov, D. and Hamza, A. M. and Mann, I. R. and Milling, D. K. and Kale, Z. C. and Chadwick, R. and Kelly, T. and Danskin, D. W. and Carrano, C. S.},
  journal={Radio Science}, 
  title={Canadian high arctic ionospheric network ({CHAIN})}, 
  year={2009},
  volume={44},
  number={01},
  pages={1--10}
}

@inproceedings{dodson2001ionospheric,
  title={Ionospheric scintillation monitoring in northern {Europe}},
  author={Dodson, Alan and Moore, Terry and Aquino, Marcio HO and Waugh, Sam},
  booktitle={Proceedings of the 14th International Technical Meeting of the Satellite Division of The Institute of Navigation (ION GPS 2001)},
  pages={2490--2498},
  year={2001}
}

@incollection{vani2021monitoring,
  title={Chapter 13 - {M}onitoring ionospheric scintillations with {GNSS in South America}: scope, results, and challenges},
  author={Bruno C{\'e}sar Vani and Alison {de Oliveira Moraes} and Lucas Alves Salles and Victor Hugo {Fernandes Breder} and Mois{\'e}s {Jos{\'e} dos Santos Freitas} and Jo{\~a}o Francisco {Galera Monico} and Eurico {Rodrigues de Paula}},
  editor = {George p. Petropoulos and Prashant K. Srivastava},
  booktitle={GPS and GNSS Technology in Geosciences},
  pages={255--280},
  year={2021},
  publisher={Elsevier}
}

@techreport{itu,
 author = {ITU},
 institution = {International Telecommunication Union},
 month = {09},
 note = {},
 number = {Recommendation ITU-R P.531-16},
 title = {Ionospheric propagation data and prediction methods required for the design of satellite networks and systems},
 year = {2025}
}

@book{wheelon2003electromagnetic,
  title={Electromagnetic scintillation: volume 2, weak scattering},
  author={Wheelon, Albert D},
  year={2006},
  publisher={Cambridge University Press}
}

@article{fremouw1978early,
  author={Fremouw, E. J. and Leadabrand, R. L. and Livingston, R. C. and Cousins, M. D. and Rino, C. L. and Fair, B. C. and Long, R. A.},
  journal={Radio Science}, 
  title={Early results from the {DNA} wideband satellite experiment---complex-signal scintillation}, 
  year={1978},
  volume={13},
  number={1},
  pages={167--187}
}

@techreport{FccStarlink,
 author = {FCC},
 institution = {Federal Communications Commission},
 month = {11},
 note = {},
 number = {DA 24-1193},
 title = {},
 year = {2024}
}

@techreport{FccAST,
 author = {FCC},
 institution = {Federal Communications Commission},
 month = {9},
 note = {},
 number = {DA 25-815},
 title = {},
 year = {2025}
}

@techreport{3gppV2,
 author = {3GPP},
 institution = {3rd Generation Partnership Project (3GPP)},
 month = {01},
 note = {Version 17.6.0},
 number = {38.108},
 title = {Satellite Access Node radio transmission and reception},
 year = {2024}
}

@article{darya2024multi,
title = {Multi-instrument analysis of {L-band} amplitude scintillation observed over the eastern {Arabian Peninsula}},
journal = {Advances in Space Research},
volume = {74},
number = {4},
pages = {1856--1867},
year = {2024},
author = {Abdollah Masoud Darya and Muhammad Mubasshir Shaikh and Grzegorz Nykiel and Essam Ghamry and Ilias Fernini},
}

@article{jha2024predicting,
year = {2024},
month = {Feb},
publisher = {The American Astronomical Society},
volume = {962},
number = {1},
pages = {L15},
author = {Jha, Bibhuti Kumar and Upton, Lisa A.},
title = {Predicting the Timing of the Solar Cycle 25 Polar Field Reversal},
journal = {The Astrophysical Journal Letters},
}

@article{caton2004gps,
  author={Caton, R. G. and McNeil, W. J. and Groves, K. M. and Basu, S.},
  journal={Radio Science}, 
  title={{GPS proxy model for real-time UHF satellite communications scintillation maps from the scintillation network decision aid (SCINDA)}}, 
  year={2004},
  volume={39},
  number={1},
  pages={1--8}
}

@article{garcia2025direct,
  author={Garcia-Cabeza, Jorge and Albert-Smet, Javier and Frias, Zoraida and Mendo, Luis and Azcoitia, Santiago Andrés and Yraola, Eduardo},
  journal={IEEE Communications Magazine}, 
  title={Direct-to-Cell: A First Look into {Starlink's} Direct Satellite-to-Device Radio Access Network through Crowdsourced Measurements}, 
  year={2026},
  volume={64},
  number={6},
  pages={52-58}}

@article{yue2014space,
author = {Yue, Xinan and Schreiner, William S. and Pedatella, Nicholas and Anthes, Richard A. and Mannucci, Anthony J. and Straus, Paul R. and Liu, Jann-Yenq},
title = {Space Weather Observations by {GNSS} Radio Occultation: From {FORMOSAT-3/COSMIC to FORMOSAT-7/COSMIC-2}},
journal = {Space Weather},
volume = {12},
number = {11},
pages = {616--621},
year = {2014}
}

@article{andrews20246g,
  author={Andrews, Jeffrey G. and Humphreys, Todd E. and Ji, Tingfang},
  journal={IEEE BITS the Information Theory Magazine}, 
  title={{6G} Takes Shape}, 
  year={2024},
  volume={4},
  number={1},
  pages={2--24}
}

@article{russell1973semiannual,
author = {Russell, C. T. and McPherron, R. L.},
title = {Semiannual variation of geomagnetic activity},
journal = {Journal of Geophysical Research (1896-1977)},
volume = {78},
number = {1},
pages = {92-108},
year = {1973}
}

@article{buschmann2024statistical,
  title={Statistical studies of plasma structuring in the auroral ionosphere by the Swarm satellites},
  author={Buschmann, Lisa Marie and Clausen, LBN and Spicher, Andres and Ivarsen, Magnus Fagernes and Miloch, Wojciech Jacek},
  journal={Journal of Geophysical Research: Space Physics},
  volume={129},
  number={2},
  pages={e2023JA032097},
  year={2024},
  publisher={Wiley Online Library}
}

@article{dunn2024temporal,
  title={Temporal variability of equatorial ionization anomaly crest locations extracted from global ionospheric maps},
  author={Dunn, Corina and Meng, Xing and Verkhoglyadova, Olga P},
  journal={Space Weather},
  volume={22},
  number={5},
  pages={e2023SW003737},
  year={2024},
  publisher={Wiley Online Library}
}

@article{Zhao2026Global,
title = {Global occurrence characteristics of ionospheric amplitude scintillation throughout a solar cycle based on {COSMIC} radio occultation observations: a case study of Solar Cycle 24},
journal = {Advances in Space Research},
volume = {77},
number = {12},
pages = {12378-12393},
year = {2026},
issn = {0273-1177},
author = {Dongsheng Zhao and Xiaoting Lai and Kefei Zhang and Craig M. Hancock and Zhongchao Shi and Longjiang Li and Wang Li and Peng Sun and Hao Liu}
}

@article{Xu2026Multi,
title = {Multi–frequency {(UHF, L, S)} scintillation at low latitudes during solar minimum: Scale–dependent irregularity response from multi–instrument observations},
journal = {Advances in Space Research},
year = {2026},
issn = {0273-1177},
author = {Zhaohui Xu and Gang Chen and Yaxian Li and Min Zhang and Zhengwen Xu and Haisheng Zhao and Shouzhi Xie and Yuanlin Jia and Chunxiao Yan and Wanlin Gong}
}

@article {Sun2026Ionospheric,
author = {Sun, Andrew K. and Morton, Y. Jade and Rino,, Charles and Lee, Jiyun},
title = {Ionospheric Scintillation Effects on {LEO}-Transmitted Signals Across Multiple Frequency Bands},
volume = {73},
number = {1},
elocation-id = {navi.772},
year = {2026},
publisher = {Institute of Navigation},
issn = {0028-1522},
journal = {NAVIGATION: Journal of the Institute of Navigation}
}

@article{Shaikh2021Daytime,
title = {Daytime {GNSS} scintillation due to {Es} over Arabian Peninsula during low solar activity},
journal = {Results in Physics},
volume = {20},
pages = {103761},
year = {2021},
issn = {2211-3797},
author = {Muhammad Mubasshir Shaikh and Govardan Gopakumar and Abdelrahman Hussein and Anton Kashcheyev and Ilias Fernini}
}

@article{Shaikh2020Occurrence,
title = {{Occurrence of pre-sunset L-band scintillation due to strong presence of sporadic-E over Arabian Peninsula}},
journal = {Advances in Space Research},
volume = {65},
number = {10},
pages = {2412-2423},
year = {2020},
issn = {0273-1177},
author = {M.M. Shaikh and I. Fernini and G. Gopakumar and N.M. Alameri}
}

@article{Shaikh2024Characteristics,
title = {Characteristics of IDL and Es layers and the impact of increasing solar activity on their descent at the Arabian Peninsula},
journal = {Advances in Space Research},
volume = {73},
number = {5},
pages = {2404-2417},
year = {2024},
issn = {0273-1177},
author = {Muhammad Mubasshir Shaikh and Manar Anwer Khaleel Abusirdaneh and Sultan Suhail Halawa and Ilias Fernini}
}

\end{document}